\documentclass{aa}

\usepackage{graphicx}
\usepackage{txfonts}
\usepackage{hyperref}
\usepackage{booktabs}
\usepackage{bm}
\usepackage[normalem]{ulem}
\usepackage[dvipsnames]{xcolor}
\usepackage{nicefrac}
\usepackage{xspace}

\usepackage{mathtools}
\usepackage{hyperref}
\hypersetup{
  colorlinks,
  citecolor=blue,
  linkcolor=blue,
  urlcolor=blue}
\usepackage{amsmath}
\usepackage{yhmath}
\usepackage{multirow}
\usepackage{array}

\newcommand{\Msun}{$\mathrm{M}_\odot$\xspace}
\newcommand{\Rsun}{$\mathrm{R}_\odot$}

\newcommand{\appropto}{\mathrel{\vcenter{
  \offinterlineskip\halign{\hfil$##$\cr
    \propto\cr\noalign{\kern2pt}\sim\cr\noalign{\kern-2pt}}}}}

\begin{document}

\title{Thermonuclear explosions as Type II supernovae}
\titlerunning{Thermonuclear Type II SNe}


\author{
    Alexandra Kozyreva\inst{1},
    Javier Mor\'an-Fraile\inst{1},
    Alexander Holas\inst{1},
    Vincent A.\ Bronner\inst{1},
    Friedrich K.\ R{\"o}pke\inst{1,2},\\
    Nikolay Pavlyuk\inst{3},
    Alexey Mironov\inst{3}\and
    Dmitry Tsvetkov\inst{3}
}
\authorrunning{A. Kozyreva et al.}
\institute{
    Heidelberger Institut f{\"u}r Theoretische Studien,
    Schloss-Wolfsbrunnenweg 35, D-69118 Heidelberg, Germany\\
    \email{sasha.kozyreva@gmail.com}
    \and
    Zentrum f\"ur Astronomie der Universit\"at Heidelberg, Institut f\"ur
    Theoretische Astrophysik, Philosophenweg 12, D-69120 Heidelberg, Germany
    \and
    M.V. Lomonosov Moscow State University, Sternberg Astronomical Institute, 119234, Moscow, Russia
    }

\date{Received; accepted }

\abstract{
We consider a binary stellar system, in which a low-mass, of 0.6~\Msun{}, carbon-oxygen white dwarf (WD) mergers with a degenerate helium core of 0.4~\Msun{} of a red giant. We analyse the outcome of a merger within a common envelope (CE).
We predict the observational properties of the resulting transient. We find that the double detonation of the WD, being a pure thermonuclear explosion and embedded into the hydrogen-rich CE, has a light curve with the distinct plateau shape, i.e. looks like a supernova (SN) Type\,IIP,  with a duration of about 40~days. We find five observed SNe~IIP: SN~2004dy, SN~2005af, SN~2005hd, SN~2007aa, and SN~2008bu, that match the $V$-band light curve of our models. Hence, we show that a thermonuclear explosion within a CE might be mistakenly identified as a SN~IIP, which are believed to be an outcome of a core-collapse neutrino-driven explosion of a massive star. We discuss a number of diagnostics, that may help to distinguish this kind of a thermonuclear explosion from a core-collapse SN.
}


\maketitle


\section[Introduction]{Introduction}
\label{sect:intro}

Detonations of sub-Chandrasekhar mass white dwarfs (WDs) in binary systems are a promising scenario for explaining Type Ia supernovae \citep[SNe,][]{2023RAA....23h2001L}. A merger of two WDs could cause a double detonation, in which one WD or both WDs explode \citep{2022MNRAS.517.5260P}. 

Here, we consider a stellar binary system, consisting of a WD which is a result of evolution of a star with initial mass of about 3~\Msun{}, and a companion star with an initial mass of 2~\Msun{}. The companion reaches the end of core hydrogen burning\footnote{More specifically, after the end of core hydrogen burning of the companion, because some time is required to ascend the red giant branch.} by the time the primary star forms a carbon-oxygen (CO) WD of about 0.6~\Msun{}. The 2~\Msun{} star becomes a red giant (RG) with a degenerate helium core (He-core), that may, if the initial system was close enough, enter into a common-envelope (CE) phase. Because of dynamical friction and tidal interaction, the WD and the He-core of the companion orbit each other inside the CE and their orbital separation shrinks. The two possible outcomes of this CE interaction are a successful envelope ejection leaving behind a close binary system of the stellar cores or a ``CE merger'' where the energy release in the orbital decay of the core binary system is not sufficient to drive envelope ejection \citep{2021ApJ...920...86K,2023LRCA....9....2R}. In the latter case, the two cores merge inside the part of the CE that is still gravitationally bound to the cores. This is the scenario we explore here. The WD and the He-core of the RG star merge, and the WD explodes as a result of double detonation \citep[see, e.g., ][]{2007A&A...476.1133F}. This WD detonation happens inside the CE, which will have a distinct effect on the final observational properties. We note that a similar scenario was proposed by \citet{Kashi2011} and \citet{2012MNRAS.419.1695I} who, however, considered different system parameters. In their ``core degenerate explosion scenario'', a CO WD merges with the CO core of a massive asymptotic giant branch (AGB) star. This is suggested to produce an object consisting of CO material and reaching or exceeding the Chandrasekhar limit\footnote{The canonical value of the Chandrasekhar limit is $M_\mathrm{cr}=5.8\,M_\odot{}/\mu_e^{\,2}${}, where $\mu_e$ is the mean molecular weight per electron \citep{1960ApJ...132..565H}.}. The outcome is hypothesized to be an explosion similar to a SN Ia but inside the remnant of an unsuccessful CE ejection. This considerably differs from our scenario, where a rather low mass CO WD interacts with the He-core of a RG star and triggers a detonation via the double-detonation mechanism. Due to the low mass of the exploding sub-Chandrasekhar mass CO WD, very little {}$^{56}$Ni is produced \citep{Javier2023} and the event arising from a pure and isolated merger of such cores may resemble a calcium-rich transient \citep{2011ApJ...738...21W,2012ApJ...755..161K}, although the nature of these objects is still under debate \citep{2021ApJ...906...65P,2022ApJ...932...58J,2023MNRAS.526..279E}.  However, in the case considered here,  this event is buried inside hydrogen-rich CE material.

The extended giant stellar envelope in our scenario in fact represents the second CE episode for this binary,
since the primary star had to undergo the giant evolutionary phase earlier in the overall binary evolution. Two CE phases in a single system might be relatively rare, though possible, since many explosive events require two CE phases.

The observational properties of a thermonuclear explosion resulting from a merger of two compact cores inside a CE have not been studied yet in detail. We therefore aim at calculating the hydrodynamical phase of the merging episode, surround the merger product with the extended hydrogen-rich stellar atmosphere of a RG, and evolve the final configuration to predict the observational signatures of such {\textbf{a}} system. The outcome of our hydrodynamical simulations of the merging episode is reported on in a separate publication by \citet{Javier2023}. In the present study, we simulate the hydrodynamical evolution of the merger product embedded into the envelope coupled with radiation and predict the observational signatures of this kind of event.

The paper is structured as follows. In Section~\ref{sect:method}, we describe
our input models. In Section~\ref{sect:results}, we present the resulting
light curves for our models, and in Section~\ref{sect:obs}, we discuss
possible candidates for the models we calculated, and in Section~\ref{sect:discuss} how to distinguish thermonuclear explosions within the hydrogen-rich CE from Type\,II SNe. We summarise our findings in Section~\ref{sect:conclusions}.


\section[Input models]{Input models and methods}
\label{sect:method}

\begin{figure}
\centering
\includegraphics[width=0.5\textwidth]{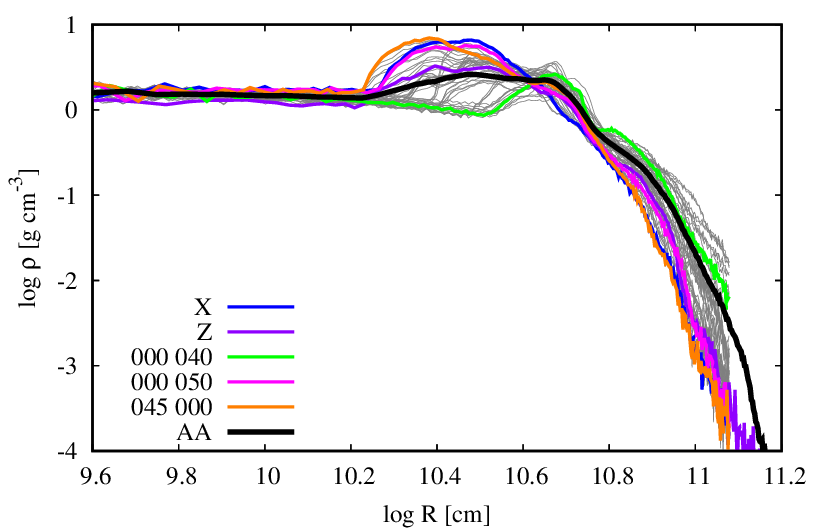}
\caption{Density profiles ($\rho$ over radius $R$) of all viewing angles (gray lines) extracted from the 3D hydrodynamics simulations of the merger of the WD and He-core with \texttt{AREPO}. The coloured curves represent the selected profiles (see Table~\ref{table:rays}) which we consider bracketing the entire set of viewing angles and used as input for radiative simulations with \texttt{STELLA}. The labels ``X'' and ``Z'' stand for the case $x$ and case $z$ in our set of selected directions.}
\label{figure:rho}
\end{figure}

\begin{figure*}
\centering
\includegraphics[width=0.9\textwidth,keepaspectratio]{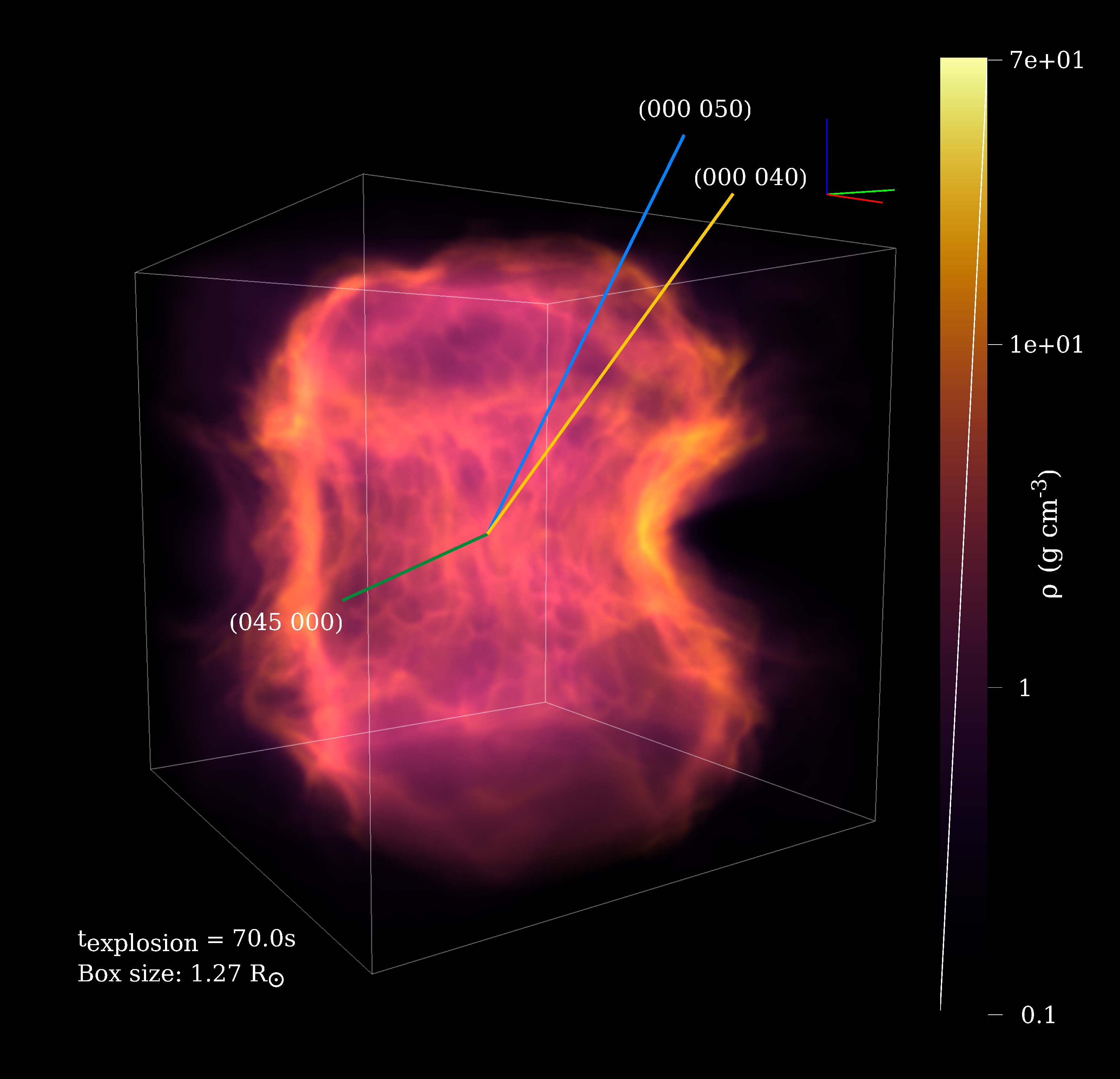}
\caption{Volume rendering of the ejecta structure of the pure merger product without the surrounding envelope, colour coded with the density. The direction of the $x$, $y$ and $z$ axes is shown in a small diagram on top of the rendering, next to the colourbar as red, green and blue lines, respectively. The coloured lines in the rendered volume illustrate three of the viewing angles used as input for the radiative-transfer simulations with
\texttt{STELLA} performed in this work.  The dark green line 045\,000 is
located in the $x-y$ plane (orbital plane of the progenitor system), the
light blue 000\,050 and yellow 000\,040 lines  are both located in the
$x-z$ plane (perpendicular to the orbital plane of the progenitor system). 
}
    \label{fig:vol-render}
\end{figure*}

In the present study, we explore a single hydrodynamical model 
of a merger of a 0.6~\Msun{} degenerate CO-core or a CO WD and a He-core of a star with an initial mass of 2~\Msun{} at the end of core hydrogen burning.

\begin{figure*}
\centering
\includegraphics[width=0.3\textwidth]{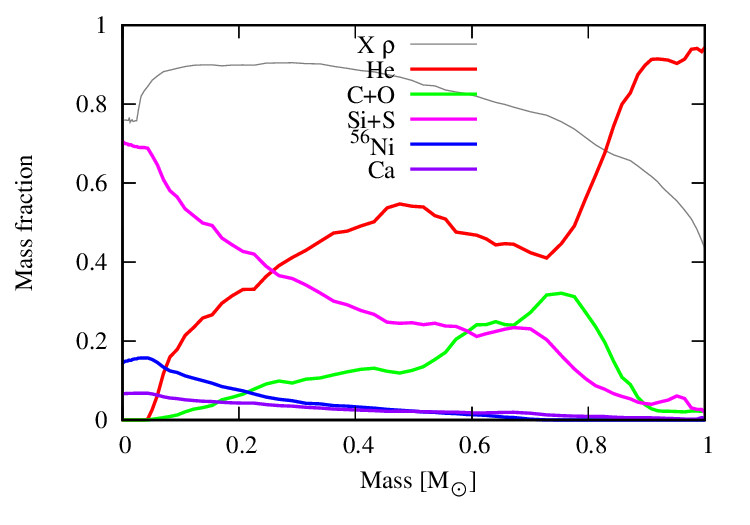}~
\hspace{-5mm}\includegraphics[width=0.3\textwidth]{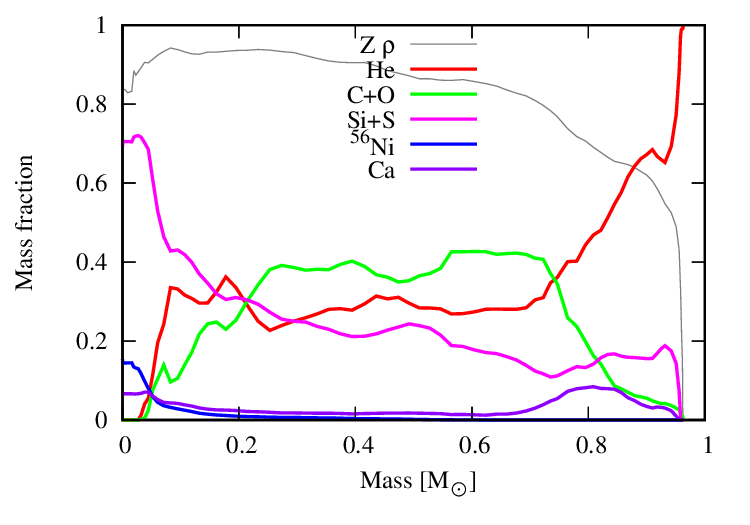}~
\hspace{-5mm}\includegraphics[width=0.3\textwidth]{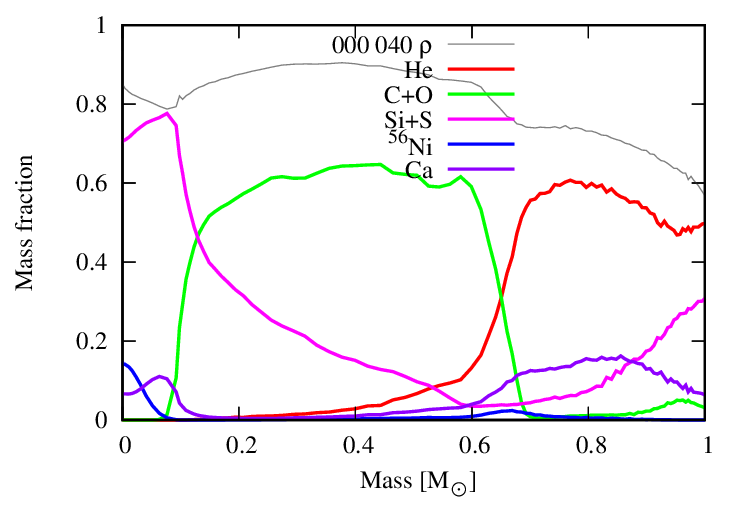}\
\includegraphics[width=0.3\textwidth]{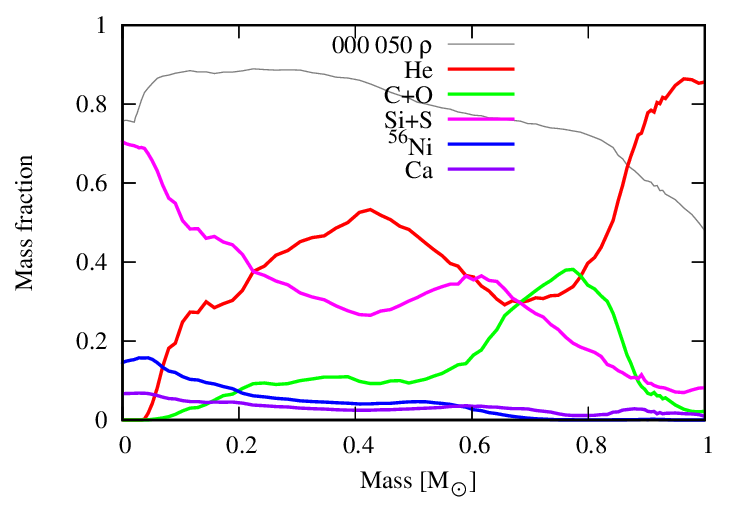}~
\hspace{-5mm}\includegraphics[width=0.3\textwidth]{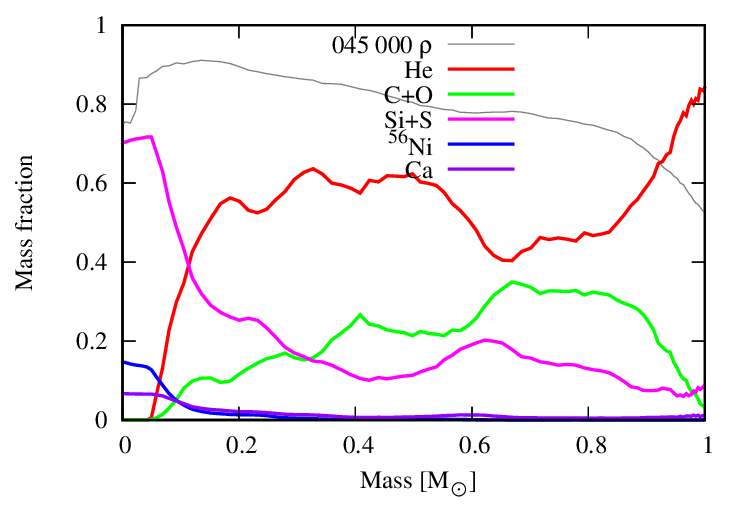}~
\hspace{-5mm}\includegraphics[width=0.3\textwidth]{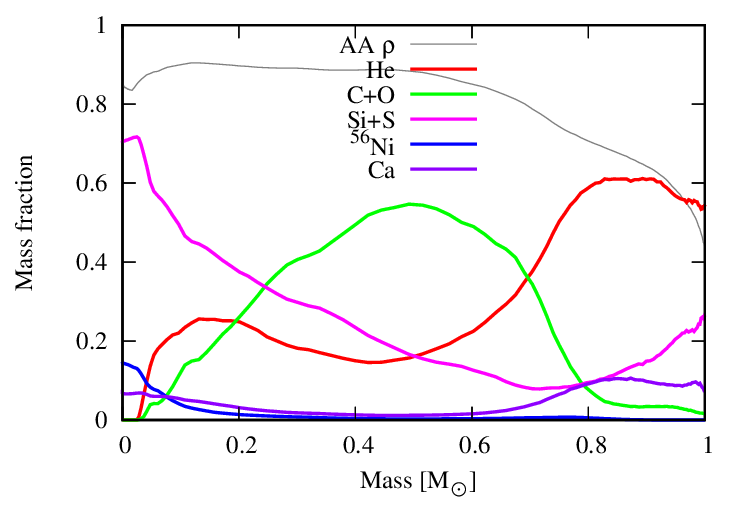}
\caption{Chemical structure of the individual rays for the pure merger. The gray curves represent scaled density profiles (see details in the text). The labels ``X'' and ``Z'' stand for the case $x$ and case $z$ in our set of selected directions.}
\label{figure:chemie}
\end{figure*}

\subsection[3D AREPO simulations of the merger]{3D AREPO simulations of the merger}
\label{subsect:arepo}

The three-dimensional (3D) simulations of the dynamical phase of the core merger are conducted with the \verb|AREPO| code
\citep{2010MNRAS.401..791S,2011MNRAS.418.1392P,2021MNRAS.503.4734P}. 
These simulations are discussed in a separate publication \citep{Javier2023}, and for details of the explosion simulation 
we refer the reader to that paper. 

We extracted numerous rays from a snapshot of the 3D simulations done with
\verb|AREPO|. Physical and chemical quantities along each ray are mass-averaged within the solid angle of $20^\circ${} around an individual ray. Among different directions we choose a few extreme directions which bracket the range of the 3D density profiles and intermediate cases. These profiles then serve as representative input models for our calculations of the light curves (LCs) and are intended to show the radiative outcome of the 
highly asymmetric geometry of the product of the merger. The same procedure is used in \citet{2023A&A...678A.170B}.

In Figure~\ref{figure:rho}, we show density profiles of
all 48 rays in gray and emphasise a few of them as thick coloured lines which we selected.  
These are: 
\begin{itemize}
\item the angle-averaged (AA, black) profile, in which all quantities are mass-weighted;
\item case $x$ (or 000\,000) along $x$-axis which points at the direction to a He-core in the orbital plane;
\item case $z$ along the $z$-axis, i.e. perpendicular to the orbital plane;
\item case 000\,040 and case 000\,050 -- both in the $x-z$ plane which is perpendicular to the orbital plane of the progenitor system;
\item case 045\,000 in the orbital plane, pointing $45^\circ${} from the $x$ axis.
\end{itemize}
The origin of the Cartesian coordinate system (zero point) is the WD centre. To illustrate the 3D geometry of the merger product, we show the ejecta structure with the selected viewing angles in Figure~\ref{fig:vol-render}. The detailed description of the ejecta is presented in \citet{Javier2023}. 

The 3D \verb|AREPO| output at $70\, \mathrm{s}$ after thermonuclear detonation of the WD, when all important nuclear reactions have ceased, was mapped into the extended atmosphere of the non-degenerate star. 
We note that nickel bubbles and, in turn, the so-called nickel-bubble effect, are likely irrelevant in our system: The total amount of radioactive nickel {}$^{56}$Ni{} contributes only 1.4\,\% of the ejecta mass of the pure merger product, and 0.5\,\% of the total ejecta mass of the merger within the envelope. Furthermore, although nickel is distributed non-isotropically, it does not form clumps, i.e. we do not expect formation of nickel bubbles later than 70~s after the WD detonation, when the merger ejecta reach homologous expansion. The nickel-rich region is confined within 0.9~\Msun{} of the ejecta and move equally at the same velocity in different directions, therefore, we exclude development of Kelvin-Helmholtz instabilities at later time. No further non-radial mixing of the ejecta material is expected too, although long-term simulations are required to confirm this assumption. 
We emphasise that we aim to simulate the merger of the WD and the He-core inside the CE, which is not ejected during the CE phase. We explain the choice of the stellar atmosphere models in Section~\ref{subsect:mesa}.

In Figure~\ref{figure:chemie}, we show the chemical structure of the merger product without envelope for the selected rays and the AA case. The gray curves in each subplot represent the density profile (scaled for illustrative purposes; for the actual physical density profiles see Figure~\ref{figure:rho}) to indicate the mass content of given species. For example, the helium fraction is relatively high in the outer part of the $x$-ray, i.e.\ along the axis connecting the WD and the He-core.
Helium is less abundant in the outer region in the directions 000\,040 and 045\,000. The thickness of the helium layer in combination with the {}$^{56}$Ni content 
influences the rise part of the LC, i.e. the lower the helium mass, the shorter the rising time. If $^{56}$Ni is present in the same region, the rise time of the LC is even shorter \citep{2013ApJ...769...67P}, as $\gamma$-ray photons produced via radioactive decay reach the photosphere earlier, causing an increase in luminosity.

In Table~\ref{table:rays}, we list the helium mass in each ray and in the AA
case for the merger product without envelope. For the mass of helium in the ejecta, we distinguish between the total mass, the mass within the inner 0.6~\Msun{} of the merger, which is associated with the WD itself, and the mass in the outer ejecta of the merger. The division is set to help in interpreting the LCs resulting from the pure WD--He-core detonation. It does not mean that the inner 0.6~\Msun{} consists of only CO WD, instead, it accounts for the macroscopic mixing happening during the accretion and ignition phase. The latter results in a large fraction of helium in the inner 0.6~\Msun{} in almost all directions. Note also that we list $4\pi$-equivalent values for the masses in Table~\ref{table:rays} which might be overestimated. We also include the {}$^{56}$Ni masses with the same meaning, particularly, to show the effect of the presence of radioactive nickel in the outer ejecta of the merger on the resulting LCs (see Section~\ref{subsect:lcs}).
In addition, we provide the terminal kinetic energy for each case
which corresponds to the total energy flowing in different directions.
The difference in the kinetic energy for different rays is explained by the different density distributions along the rays, as velocity profiles tend to be very similar to each other. We do not list the mass of the hydrogen-rich CE in Table~\ref{table:rays}, because the envelope is equivalent in the sense of mass and composition in all directions. We provide details about the CE in Section~\ref{subsect:mesa}.

\begin{table}
\caption{$4\,\pi-$equivalent parameters of the pure merger without an envelope taken at different rays.}
\begin{center}
\begin{tabular}{>{\centering}p{1.15cm}| p{0.45cm}p{0.45cm}p{0.65cm}| p{0.60cm}p{0.60cm}p{0.70cm}| p{0.4cm}  }
\hline \multirow{2}{*}{Ray} & \multicolumn{3}{c|}{He [\Msun{}]} & \multicolumn{3}{c|}{$^{56}$Ni [\Msun{}]} & \multirow{2}{*}{E$_{50}$} \\ 
{} & M$_{\textrm{tot}}$& M$_{\textrm{in}}$& M$_{\textrm{out}}$& M$_{\textrm{tot}}$& M$_{\textrm{in}}$& M$_{\textrm{out}}$ & {} \\  \hline 
      X&0.51&0.22&0.29&0.038&0.037&0.001&3.6\\
      Z&0.33&0.15&0.17&0.010&0.010&0.000&4.8\\
000\,040&0.25&0.02&0.23&0.009&0.006&0.003&7.6\\
000 050&0.46&0.21&0.25&0.043&0.041&0.002&4.4\\
045 000&0.56&0.29&0.27&0.014&0.014&0.000&4.1\\
     AA&0.32&0.11&0.21&0.014&0.013&0.001&5.9\\ \hline
\end{tabular}
\label{table:rays}
\end{center}
{{\textbf{Notes:}}Total mass of helium and radioactive nickel {}$^{56}$Ni in \Msun{} 
(total mass $M_\mathrm{tot}$, mass within inner 0.6~\Msun{} and
in the outer 0.4~\Msun{}), terminal kinetic energy as a representative
of explosion energy E$_{50}$ in the units of $10^{\,50}$~erg.}
\end{table}

\subsection[1D MESA simulations of the secondary star]{1D MESA simulations of the secondary star}
\label{subsect:mesa}

The goal of the study is to investigate the observational implications of a thermonuclear explosion arising from the core merger interacting with CE material. Because we consider a system where CE ejection has not successfully completed and the cores merge instead of forming a close binary system of compact cores, we have to embed the explosion model presented by \citet{Javier2023} in a model for the CE material surrounding it. For simplicity, we consider two cases: the unperturbed envelope of the RG star in hydrostatic equilibrium and a perturbed RG envelope that results from the framework of the parametrised one-dimensional CE interaction model described by \citet{Bronner2023}.

As discussed in \citet{Javier2023}, a successful detonation of the primary 0.6~\Msun{} CO WD seems to only be possible for He-cores (or He WDs) in a narrow range around 0.4~\Msun{}. This is due to the He ignition mechanism that ignites the double detonation, relying on the He-core being disrupted relatively close to the CO WD. The tidal forces acting on a significantly less massive core result on a disruption further away from the surface of the CO WD, and the conditions for He ignition never being reached.
The corresponding ZAMS mass of a companion star that would lead to a successful double detonations is about 2~\Msun{} to 2.5~\Msun{}, assuming solar metallicity. Therefore, we modeled a 2~\Msun{} star with metallicity $Z=0.02$ as a representative case. We calculated stellar evolution with \texttt{MESA} version 12778 \citep{2011ApJS..192....3P,2013ApJS..208....4P,2015ApJS..220...15P,2018ApJS..234...34P,2019ApJS..243...10P}. 

The simulation started at the zero-age main-sequence and was stopped once the He-core mass reached 0.4~\Msun{}, with the helium core being defined by a hydrogen mass-fraction less than 0.1. This corresponds to a stellar age of $1.07 \times 10^{\,9} \, \mathrm{yr}$. We set the mixing-length parameter $\alpha_\mathrm{MLT} = 2$. For the RG winds, we use the Reimers prescription with $\eta = 0.5$ \citep{1975MSRSL...8..369R}. 
The initial solar metallicity means that the metal content is about 0.02. MESA distributes
metal content between $^{12}$C, $^{14}$N, $^{16}$O, $^{20}$Ne, $^{24}$Mg with a
ratio of 3:1:9:2:4. By the time when the star reaches RG phase the ratio between these species is 2:2:9:2:4,  which is different to the initial ratio because the first dredge-up episode takes place. We note that the MESA profiles
have zero stable iron abundance. This is due to the choice of the small nuclear network \texttt{basic}, which is justified, since the calculated evolution does not proceed to the late stages of stellar evolution. In the
following radiative transfer simulations the presence of iron plays significant
role, which we will discuss in Section~\ref{subsect:Zdepend}.

At the end of the simulation, the star has a radius of 75~\Rsun{} and a mass of 1.9~\Msun{}. Then, we perturbed the envelope of the secondary to mimic a CE evolution following the model described in \citet{Bronner2023}. This model integrates the orbits of the two stars in the CE by assuming a drag force \citep{2010ApJ...725.1069K} and injects the released orbital energy as heat into the envelope. We choose $a_\mathrm{ini} = 70\,\mathrm{R}_\odot${} for the initial separation and $C_\mathrm{d}=0.25$ and $C_\mathrm{h}=2.0$ for the two free parameters. The numerical values for $C_\mathrm{d}$ and $C_\mathrm{h}$ were inspired by the CE simulations of a 1~\Msun{} AGB star described in \citet{Bronner2023}. The CE simulation lasts for 77 d, during which the envelope expands to about 330~\Rsun{}. 
The calculation with the unperturbed envelope is used as a reference point to illustrate the effect of an inflated envelope around the detonating merger.
As our primarily interest lies along the perturbed case we name this kind of envelope ``env1'', while we keep unperturbed case as a reference with the name ``env2''. 

The hydrogen-rich atmosphere of the giant star with the extracted He-core has 1.5~\Msun{}. In order to append the atmosphere, we link the profiles of both cases of the envelope to selected profiles of the merger product (1~\Msun{}), so that the junction is smooth density-wise.

\subsection[1D radiative-transfer STELLA simulations]{1D radiative-transfer STELLA simulations}
\label{subsect:stella}

The profile representing the AA case and profiles in selected directions
were mapped into the 1D radiation-hydrodynamics
code \verb|STELLA| \citep{2006AandA...453..229B}\footnote{The version of STELLA used in the current study is the private and not the one
implemented in MESA \citep{2018ApJS..234...34P}}.  \verb|STELLA| is
capable of modeling hydrodynamics, including shock propagation and its
interaction with the medium, as well as the radiation field
evolution, i.e.\ it self-consistently computes the LCs, the spectral energy distribution and the resulting
broad-band magnitudes and colours.  We use the standard parameter settings, that are 
explained in many papers involving \verb|STELLA| simulations \citep[see e.g.,][]{2021AstL...47..291T,2020MNRAS.497.1619M}.  The
thermalisation parameter is set to $0.9$ as recommended by the recent study of \citet{2020MNRAS.499.4312K}.
The profiles in different directions are considered as $4\,\pi-$equivalent
spherically-symmetric models and cannot represent the entire picture of a
full 3D radiative transfer simulations as it can be done with \verb|ARTIS|, \verb|SEDONA|,
\verb|SuperNu| and other sophisticated spectral synthesis codes
\citep{2022A&A...668A.163B}, which, however, lack models for the interaction of hydrodynamics and radiation.


\section[Results]{Results and Discussion}
\label{sect:results}

\subsection[]{Light curves}
\label{subsect:lcs}

\begin{figure}
\centering
\includegraphics[width=0.5\textwidth]{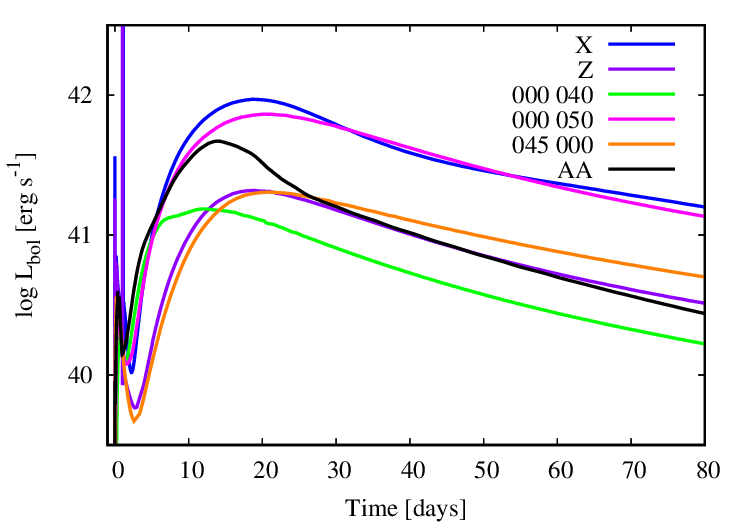}
\caption{Bolometric LCs for pure merger product without envelope for the selected rays and the AA case. The labels ``X'' and ``Z'' stand for the case $x$ and case $z$ in our set of selected directions.}
\label{figure:bol}
\end{figure}

In Figure~\ref{figure:bol}, we present bolometric LCs\footnote{The data computed and analysed for the current study are available via the link
\href{https://doi.org/10.5281/zenodo.10000753}{https://doi.org/10.5281/zenodo.10000753}.
} for the pure merger product without envelope seen from different viewing angles and the AA case. We calculate and show these LCs as a reference to demonstrate how different the LCs for the merger with envelope look. The LCs resemble LCs of SNe~Ia, i.e.\ they are powered by the radioactive decay $^{56}$Ni and $^{56}$Co. The peak luminosity of the
nickel-powered LC is expected to be connected to the mass of $^{56}$Ni
following the relation \citep{1979ApJ...230L..37A}:
\begin{equation}
L_\mathrm{\,peak} \sim M_\mathrm{\,Ni} \, \varepsilon(t_\mathrm{\,peak}) \,,
\label{equation:LNI}
\end{equation}
where $\varepsilon(t_\mathrm{\,peak})$ is the decay function of nickel and cobalt.  
The shape of the these LCs depends also on distribution of radioactive {}$^{56}$Ni
within the ejecta and the ejecta mass. The shallower the $^{56}$Ni distribution, the
earlier and the lower the main peak compared to the centrally concentrated {}$^{56}$Ni distribution.
For instance, the presence of radioactive  $^{56}$Ni in the outer layers
in the 000\,040 case shortens the rise time 
\citep{2013ApJ...769...67P,2014ApJ...784...85P}, and even produces a
bump in the early LC \citep{2017MNRAS.472.2787N,2020A&A...634A..37M}. 
We note that the $^{56}$Ni masses for individual rays listed in Table~\ref{table:rays}
are considered as $4\,\pi$-equivalent and not necessarily correspond to the
actual mass of this isotope. As discussed in Section~\ref{sect:method}, 
the rise time of the LC depends also on the mass of material lying on top of the
{}$^{56}$Ni-enriched layers, particularly, the amount of helium in the outer part of the ejecta. Helium being mixed with a small fraction of radioactive nickel {}$^{56}$Ni is non-thermally ionised and produces higher opacity, in turn, increasing the rise time \citep{2012MNRAS.424.2139D}. Thus, the 000\,040 LC has the shortest rise to the peak, as this
input model has the lowest helium content and some amount of {}$^{56}$Ni in the outer layers.
Another impact of the high degree of mixing of {}$^{56}$Ni is the redder colours of the resulting LCs \citep{2019ApJ...872..174Y}, because nickel (similar to iron and other iron-group elements) has a high line opacity, and efficiently redistributes blue flux into redder wavelengths \citep{2006ApJ...649..939K}.

In realistic 3D simulations the LC will depend on the viewing angle and should be a sum
of the photons streaming in different rays. With our 1D simulations we show
the scatter of the final LCs the observer can see from different viewing
angles, which depends on the position of the asymmetric merging
system relative to a point of view. The LCs computed with \verb|STELLA| agree reasonably well with those
simulated with 3D version of \verb|ARTIS| \citep{Javier2023}.

\begin{figure*}
\centering
\includegraphics[width=0.5\textwidth]{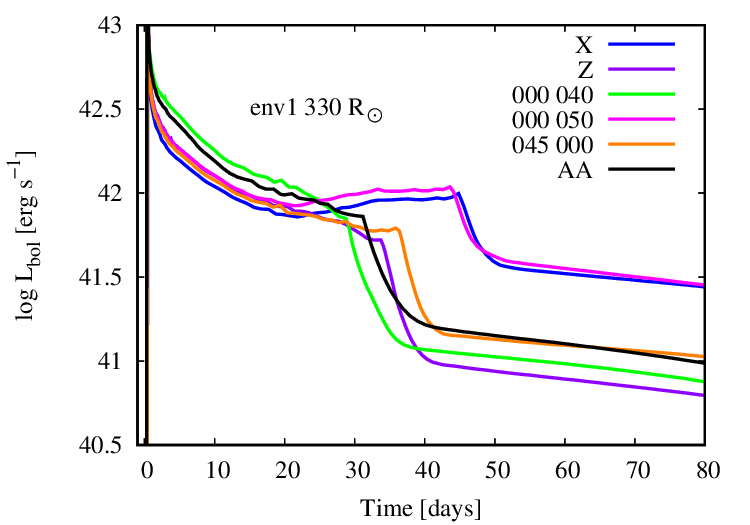}~
\includegraphics[width=0.5\textwidth]{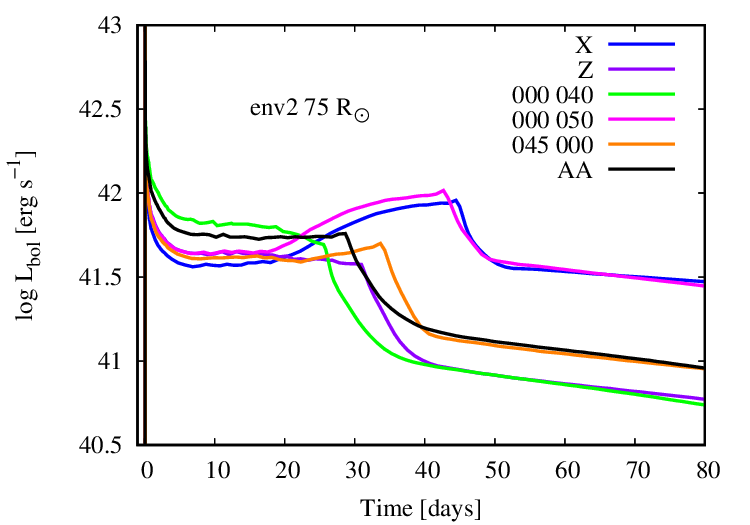}
\caption{Bolometric LCs for the selected rays and the AA case embedded into
extended (``env1'', left) and compact (``env2'', right) envelopes. The labels ``X'' and ``Z'' stand for the case $x$ and case $z$ in our set of selected directions.}
\label{figure:bolenv1env2}
\end{figure*}

When embedded in the hydrogen-rich envelope, the 
original shape of a SN Ia-like LC of the pure merger (see Fig. 3 in \citealt[][]{Javier2023}) is significantly erased,
as it is seen in Figure~\ref{figure:bolenv1env2}. 
The resulting LCs exhibit a pronounced plateau, which resembles that of a
hydrogen-rich SN~II. The luminosity on the plateau and its duration vary within
0.3~dex and depend on the effective explosion energy
according to the scaled relations of \citet{1993ApJ...414..712P}: $\log L_\mathrm{plateau}\sim 0.8 \log E_\mathrm{expl}${}, where the dependence on mass and radius is irrelevant for our case because radius and mass are the same for all
viewing angles. The duration of the plateau varies between 30~days and 45~days for
the cases of low (ray\,000\,040, ray\,045\,00, $z$-ray, and AA case) and high
($x$-ray and ray\,000\,050) $^{56}$Ni mass (see Table~\ref{table:rays}), respectively.
The LCs for the $x$-ray and the ray\,000\,050 show the delayed $^{56}$Ni-powered
contribution to the overall LCs, extending the plateau by ten more days.
The influence of {}$^{56}$Ni in extending and lifting the plateau is also present in usual SNe~IIP \citep{2009ApJ...703.2205K,2019ApJ...879....3G,2019MNRAS.483.1211K}.
The larger radius of the extended envelope ``env1'', 330~\Rsun{}, causes 
shallower decline in LCs after shock breakout (see Figure~\ref{figure:bolenv1env2}), according to the time $t_\mathrm{rec}\sim R^{\,0.76}$ when recombination settles in the outer ejecta \citep{Shussman2016}. In contrast, the LCs of the merger in the
compact envelope ``env2'' (75~\Rsun{}) drop quickly after shock breakout.

As seen in Figure~\ref{figure:bolenv1env2}, there are two extreme cases
among all considered models in our study. Therefore, we pick the $x$-ray and the AA
case as boundary examples to show the broad band LCs in the $U$, $B$, $V$, and $R$
bands in Figure~\ref{figure:UBVRadd} in Appendix~\ref{appendix:append1}.
Interestingly, the luminosity of the low-energy (0.6~foe) transient
considered in our study is similar to plateau luminosity found for average SNe~IIP.
Particularly, the models including envelope ``env1'', i.e.
extending to 330~\Rsun{}, have up to $-17$~mags in $U$-band, $-16.3$~mags in
$B$-band, $-16.4$~mags in $V$-band, and $-16.3$~mags in $R$-band, which is seen in
SN\,1999em \citep{2003MNRAS.338..939E}. However, the distinct difference of
our transients is the very short plateau, lasting for only 30\,--\,45~days, while a usual duration of the plateau in SNe~IIP is 100~days to 150~days. 

\subsection[Dependence on metallicity]{Dependence on metallicity}
\label{subsect:Zdepend}

\begin{figure*}
\centering
\includegraphics[width=0.5\textwidth]{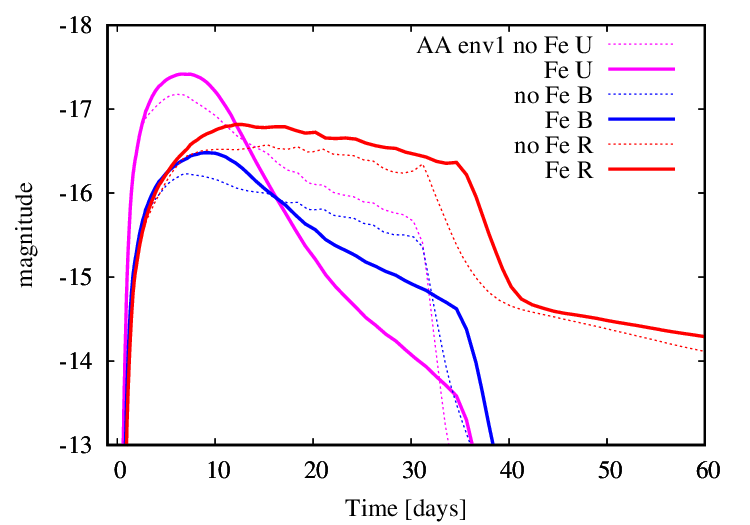}~
\includegraphics[width=0.5\textwidth]{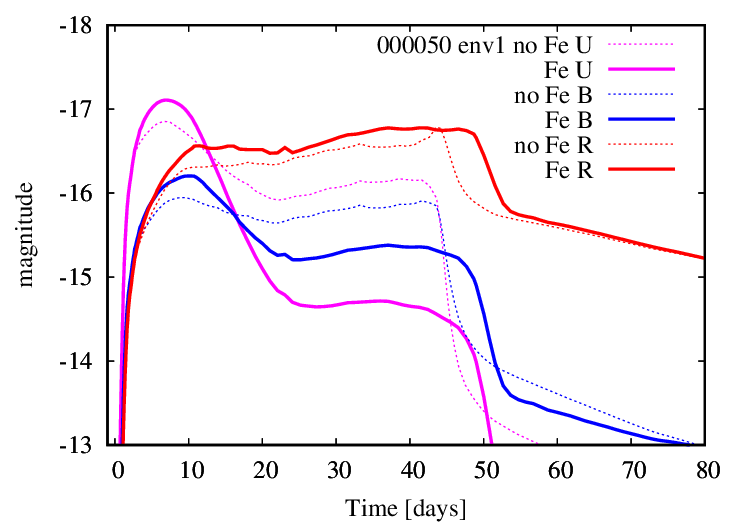}
\caption{Metallicity dependence for the extended envelope: 
$U$ (magenta), $B$ (blue), $R$ (red) LCs showing the difference
between the case of iron-free envelope (dashed) and iron-polluted envelope (solid) for
the AA case (left) and the ray 000\,050 (right).}
\label{figure:Zdepend}
\end{figure*}

As described in Section~\ref{subsect:mesa}, the chemical composition of the modelled stellar envelopes does not include
iron, as described in Section~\ref{sect:method}. However, iron plays an important role in radiative transfer simulations, in particular for shaping the X-ray, $U$- and
$V$-band LCs. Therefore, we calculated an additional set of models for two
cases, AA and ray 000\,050 for both cases of the envelope, extended (``env1'') and
compact (``env2''), with solar metallicity abundance of iron \citep[$X(\mathrm{Fe})=1.46\times10^{\,-3}$,][]{2003ApJ...591.1220L}.
The resulting comparison plots are presented in Figure~\ref{figure:Zdepend}
and in Figure~\ref{figure:ZdependAdd} in Appendix~\ref{appendix:append2}.
We note that the main influence of iron in the hydrogen-rich envelope is line blanketing in the bluer wavelengths, while it does not affect redder wavelengths ($V$-band and beyond). The bolometric LCs are also not heavily affected by
iron similar to $V$ and $R$-bands, since flux in these bands mostly contributes to overall luminosity, except the length of the plateau. The plateau is 5~days longer in the case of non-zero abundance of iron, since iron contributes to the overall
opacity, making the optical depth and diffusion time longer for the
hydrogen-rich envelope. The latter can be seen, for example, in the $R$-band
LCs. The $U$-band LCs decline faster in the case of non-zero iron abundance in the envelope in comparison to the case of the iron-free envelope, and the luminosity at the end of the plateau is more than 1~mag lower for the iron-enriched case.


\section[Looking for the possible observed candidates]{Looking for the possible observed candidates}
\label{sect:obs}

\begin{figure*}
\centering
\includegraphics[width=\textwidth]{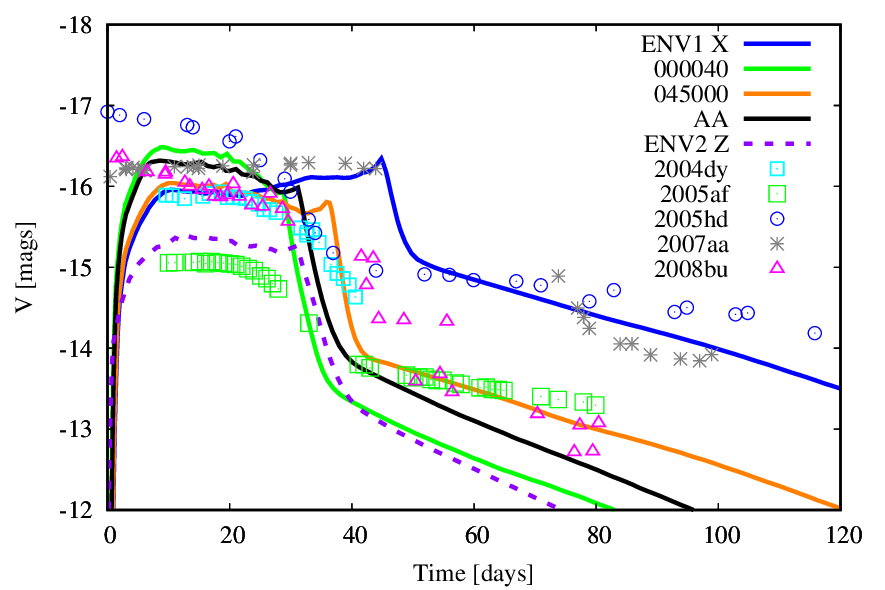}
\caption{$V$-band LCs for the models in the study and SN~2004dy, SN~2005af, SN~2005hd,
SN~2007aa, SN~2008bu.}
\label{figure:obsV}
\end{figure*}

While looking for the possible observed candidates of our modeled events, we apply the following
criteria: the luminosity on the plateau should be a little bit lower than usual
for SNe~IIP ($-16$~mags{}) and -- the most important criterion -- the duration of the plateau should not exceed 40~days. These criteria are strict and difficult to match.
However, we find a few possible SNe with similar properties in a few
observational sets such as \citet{2014ApJ...786...67A}\footnote{The list of Transient surveys used can be found in the publication.} and \citet{2022A&A...660A..40M}\footnote{Carnegie Supernova Project-I.}. 
We limit our comparison to the $V$-band, because in many cases data in other broad bands are unavailable and because detailed comparison between models and observations is beyond the scope of the current study. 
The observational data are taken from the Open Supernova Catalogue \citep{2017ApJ...835...64G}\footnote{ \url{https://github.com/astrocatalogs/supernovae}{}, \qquad \url{https://sne.space/}.}.

In total, we find five examples among the observed SNe~II
with low-to-intermedium luminosity on the plateau and short plateau duration.
The best matching synthetic LCs are from the following models:
$x$-ray, 040\,000 and 045\,000 rays, and AA case in the extended
envelope; and $z$-ray, embedded in the compact envelope.
In Figure~\ref{figure:obsV}, we compare the $V$-band magnitude for our selected rays and the AA case to data from SN~2004dy, SN~2005af, SN~2005hd, SN~2007aa, and SN~2008bu. 
The cases of SN~2004dy, SN~2005hd, SN~2007aa, and SN~2008bu display $V$-band LCs similar to our models in the extended envelope, while SN~2005af\footnote{We note, that the explosion epoch for SN\,2005af is uncertain. Hence, \citet{2005IAUC.8484....2F} report that the explosion happened ``perhaps a month before'' their acquired spectrum. Furthermore, \citet{2006ApJ...651L.117K} claim that ``the explosion epoch is probably uncertain by up to a few weeks''. Therefore, 2005af should be considered as a candidate for our merger with some caution.} might be considered close to the $z$-ray in the compact envelope. As mentioned above, the detailed analysis of our models and the candidates is beyond of the scope of the current study, we find some additional pieces of data for the candidates and discuss this in Section~\ref{subsect:spectra}.

As we note above in Section~\ref{subsect:lcs}, the luminosity on the plateau
is not a strong criterion for looking for possible candidates. Our models
within the extended envelope have $-16.3$~mags to $-17$~mags in $U$, $B$,
$V$, and $R$ broad bands, which is close to normal values for SNe~IIP, even
though slightly lower. The so-called low-luminosity SNe~IIP have even
lower luminosities on the plateau,  \citep[about $-14$~mags, e.g., SN~2020cxd and
SN~2005cs, see][]{2009MNRAS.394.2266P,2021AandA...655A..90Y,2022MNRAS.513.4983V},
but the duration of the plateau is 150~days, which is mainly the signature
of a relatively low-energy explosion. Therefore, the study by
\citet{2022MNRAS.514.4173K} shows that these SNe can be
easily explained by neutrino-driven core-collapse explosions of low-mass
massive stars in the initial mass range around 9~\Msun{}.
The very low-luminosity event SN~1997D with a $V$--band magnitude of
$-14.65$~mags suffers from a significant uncertainty in the explosion epoch, 
although the best estimate based on the spectral synthesis and colour
evolution shows that plateau duration of SN~1997D is longer than 50~days
\citep{1998ApJ...498L.129T,2000A&A...354..557C,2001MNRAS.322..361B,2003MNRAS.338..711Z}, i.e. it might not be a good match for our models.
The recently reported luminous short-plateau events SNe~2006Y, 2006ai, and 2016eg have 
plateaus of 50 to 70~days, and their plateau luminosity is relatively
high, $-17$ to $-17.5$~mags \citep{2021ApJ...913...55H}, which is more
likely to be explained by relatively high for CCSNe explosion energies \citep{1993ApJ...414..712P}.

We conclude that our simulated merger product being surrounded by an
extended stellar atmosphere of a giant star can be seen as a
very short-plateau SN~IIP. However, the host galaxy of this SN should contain an old
stellar population, or the SN location should be associated with the old
population regions of a galaxy. The rate of this kind of explosion can reach up
to a few percent that of all SNe~IIP, if looking into the observations by
\citet{2014ApJ...786...67A}. Nevertheless, the detection of transients matching the predictions of the models in our study is complicated because of their short 30\,--\,45 day plateau, i.e. their detectability is lower in comparison to the 100~day lasting SNe~IIP, as some of these short-living transients can be missed. 
In reality, the probability of the possible candidates for our models, specifically those within a compact stellar atmosphere, is even lower, because of relatively low luminosity, which depends on the total amount of detected SNe~IIP.
Next generation transient facilities such as James Webb Space Telescope (JWST), Euclid Telescope, Roman Space Telescopes, and Rubin Observatory (LSST) will provide a
deeper observations and cover a larger volume of the Universe. As a
result, a larger number of low-luminosity transients will be detected potentially including events that can be explained by our models.


\section[How to distinguish thermonuclear Type II SNe]{How to distinguish thermonuclear Type II SNe}
\label{sect:discuss}

In the previous sections, we have shown that the LC resulting from the merger of a low-mass WD and a He-core of a RG star happening inside the CE is similar to a SN~II-P, although having some distinct properties. The mechanism providing the energy deep inside the hydrogen-rich envelope, however, is a thermonuclear explosion instead of a gravitational collapse of a stellar core. This leads to the question of how to distinguish this scenario from the common gravitational-collapse induced SNe~II. In the following, we discuss how electromagnetic observables can break this degeneracy, and how non-electromagnetic observables can provide a solid distinction between the two explosion scenarios.

\subsection{Spectra}
\label{subsect:spectra}

\begin{figure}
\centering
\includegraphics[width=0.49\textwidth]{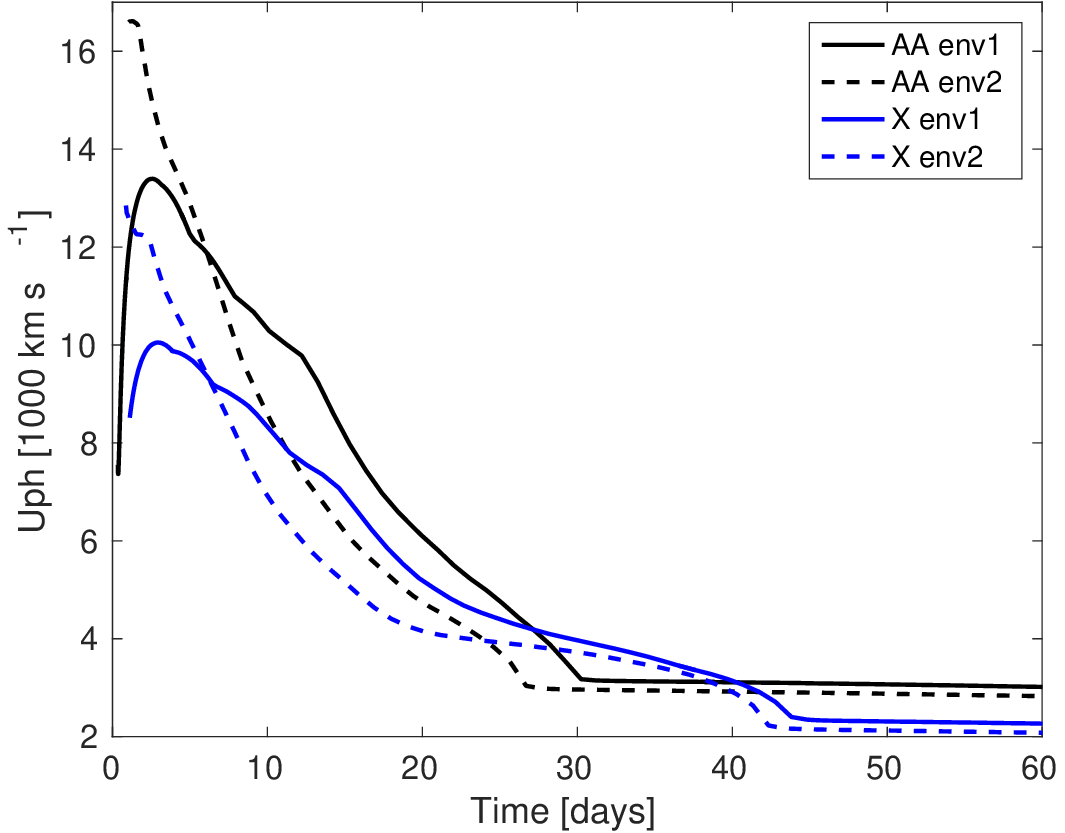}
\caption{
Photospheric velocity evolution for the AA case (black) and the
$x$-ray (blue) in the extended ``env1'' (solid) and compact ``env2'' (dashed) envelopes.}
\label{figure:uph}
\end{figure}

In the  present study, we carry out radiative transfer simulations with the
multi-group code \verb|STELLA|, which does not allow us to calculate spectra. However, we attempt
to understand potential spectral signatures of the merger of a CO WD and a He-core within a CE. 

\begin{figure}
\centering
\includegraphics[width=0.49\textwidth]{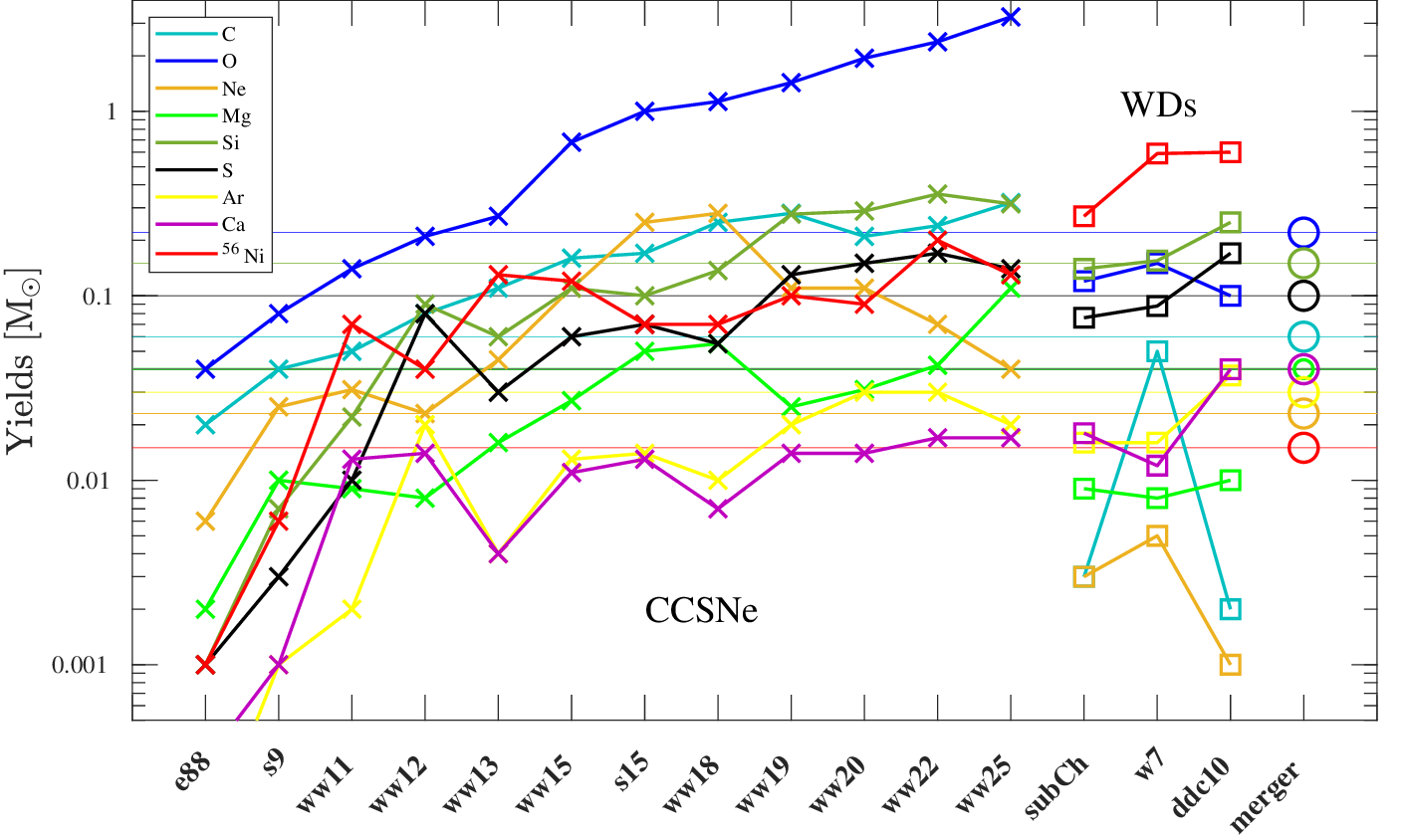}
\caption{Total yields of carbon, oxygen, neon, magnesium, silicon, sulphur, argon, calcium, and radioactive nickel {}$^{56}$Ni. The symbol ``$\times$'' marks the yields for core-collapse explosions \citep{1995ApJS..101..181W}, whereas models of thermonuclear WD explosions are marked with ``$\square$''. Circles present yields for our model. Horizontal lines with the colour corresponding to the selected species are inserted for convenience to compare the merger yields with other explosions.}
\label{figure:yields}
\end{figure}

During the first 40~days the photosphere propagates through the extended
hydrogen-rich part of the ejecta, and it is unambiguous that  spectra of the merger during this phase will resemble those of SNe~IIP, i.e. will exhibit pronounced hydrogen lines.
Figure~\ref{figure:uph} shows the photospheric velocity evolution for the merger. We estimate this quantity as the velocity of a mass-shell in which the integrated optical depth in the $B$-band equals $\tau=\nicefrac{2}{3}$.
Velocities evolve quickly from 12,000\,km\,s$^{\,-1}$ to 3,000\,km\,s$^{\,-1}$ over the first 40~days, i.e. the photospheric velocities estimated during this
epoch are typical for SNe~IIP. 
The decline in the velocity evolution is sharp, and on day~40 after the onset of the explosion the photosphere reaches a layer which moves at 3,000\,km\,s$^{\,-1}$ and slower.  The photospheric phase turns to the nebular phase, i.e.\ the spectra become dominated by distinct lines, which 
will resemble the outcome of an explosion of a massive star, if looking at the line width, because of a similar explosion energy of 0.6~foe \citep{2016ApJ...818..124E,2020ApJ...890...51E}\footnote{More precisely, the width of lines corresponds to the quantity $\sqrt{E_\mathrm{expl}/M_\mathrm{ej}}$.}.

After day~40, the photosphere moves into the inner part of the ejecta, i.e.\ the actual merger product, and the inner mechanism of the SN event becomes revealed to electromagnetic observation. We expect that, at this time, there will be a transition to the nebular phase for our event, i.e. earlier than in Type\,II SNe (day~150), since the photospheric phase is over. The spectra should show lines of He, O, Ca, Si and other heavy elements on top of the typical spectra of Type~II SNe, namely, hydrogen lines, because the presence of hydrogen-rich envelope cannot be hidden and will contribute to the spectra even after the average photosphere move to the inner ejecta \citep[e.g., ][]{2012A&A...546A..28J}. An attempt in modelling a thermonuclear explosion of a Chandrasekhar-mass WD (with 0.3~\Msun{} of {}$^{56}$Ni) within a circumstellar material, i.e.\ CE  with the total mass of 13~\Msun{}, is proposed in \citet{2020Sci...367..415J}. The nebular spectra of their transient were a mixture of spectra of a WD explosion and strong H$\alpha$ line. 

To make a more educated guess about the resulting spectra, we analyse the total yields of our merger in comparison to yields for core-collapse explosions and three models of thermonuclear WD explosions. The yields for CCSNe are: electron-capture SN ``e88'' with the initial mass of of 8.8~\Msun{} \citep{2021MNRAS.503..797K}, ``s9'' (9~\Msun{}) and ``s15'' (15~\Msun{}) progenitors from \citet{2016ApJ...821...38S}, and the set of models ``ww11...ww25'' \citep[][their models s11A...s25A, correspondingly]{1995ApJS..101..181W}. Thermonuclear explosions include: an explosion of a sub-Chandrasekhar mass WD \citep[Model~1 of][]{1994ApJ...423..371W}, the W7 model of an explosion of a Chandrasekhar-mass WD \citep{1984ApJ...286..644N}, and the Chandrasekhar-mass WD explosion model DDC10 \citep{2013MNRAS.429.2127B}.
We show yields of the selected species in Figure~\ref{figure:yields}. The yields of carbon, oxygen, and {}$^{56}$Ni in our merger model are closer to the yields of core-collapse events in low-mass star range (9 to 12~\Msun{}). The yields in neon, magnesium, silicon, sulphur, and argon yields are similar to both core-collapse events in stars above 15~\Msun{} and thermonuclear explosions of WDs, while the calcium yield is clearly more compatible with the thermonuclear explosion DDC10. Therefore, spectra of the merger might contain features which can point at a core-collapse of a massive star in a entire range of CCSN progenitors, or a WD explosion at the same time. However, it is difficult to assess the strength of the spectral lines of these species in our merger in comparison to core-collapse SNe (CCSNe) and WD explosions, since ionisation states can differ significantly. Based on the nebular spectral synthesis by \citet{2020Sci...367..415J}, we guess, that spectra after about day~40  
will appear as a blend of Type~I SNe with pronounced H$\alpha$ line, which is
a diagnostic for the transients analysed in this study. For a preliminary investigation of the spectra without the hydrogen-rich envelope, see \cite{Javier2023}.
The lines of the elements resulting from the thermonuclear burning are expected\ to be narrower, in contrast to the  typical photospheric velocity of a SN~Ia about 10,000~km\,s$^{\,-1}$ during the first 60~days after the explosion.

The nebular spectra for one of our possible observed candidates, namely SN\,2005af are published by \citet{2006ApJ...651L.117K}. Among the spectral features usual for a SN\,IIP, SN\,2005af shows strong $[$Ar\,II$]$ and $[$Ni\,II$]$ lines, which are claimed in \citet{2006ApJ...651L.117K} to be stronger than in usual SNe\,IIP. Their estimates for the lower limits of the Ne, Ar, and stable Ni yields are $10^{\,-3}$~\Msun{}, $2.2\times10^{\,-3}$~\Msun{}, and $3.7\times10^{\,-3}$~\Msun{}, respectively, although 
these estimates are almost ten times lower than those for our merger model. The total mass of radioactive nickel {}$^{56}$Ni is 0.027~\Msun{}, as estimated in \citet{2006ApJ...651L.117K}, which is higher than that produced by our merger (0.014~\Msun{}). However, the detected degree of polarisation, which corresponds to 20\% asphericity of the SN ejecta \citep{2006A&A...454..827P} may favor the ``merger-CE'' origin of SN\,2005af.
The spectra for another candidate, SN\,2004dy, show a mixture of a SN\,IIP and a SN\,Ib$/$c, and therefore it was identified as peculiar Type\,II SN (IAUC~8404 and
IAUC~8409)\footnote{
{\small{\url{https://sites.astro.caltech.edu/~avishay/cccp/texts/sn2004dy.html}}}\\
\url{http://www.cbat.eps.harvard.edu/iauc/08400/08404.html}\\ \url{http://www.cbat.eps.harvard.edu/iauc/08400/08409.html}
}. It has been suggested that the presence of the strong He\,I 5876 (which is found to be much stronger than usual in SNe\,IIP) requires high helium abundance; however, no quantitative analysis of the helium mass in this SN is available in the literature. For comparison, the helium yield is 0.72~\Msun{} in our merger with CE material included, while a 15~\Msun{} CCSN progenitor yields about 5~\Msun{} of helium, although the conditions for forming lines can be different. In fact, there is an ongoing debate about the amount of hidden helium in the ejecta in different types of SNe \citep{2012MNRAS.422...70H,2021ApJ...908..150W}. We note, that helium will produce a distinct emission line, when being non-thermally excited, i.e. if it is located within one mean-free path of $\gamma$-rays from radioactive nickel {}$^{56}$Ni  \citep{2012MNRAS.424.2139D}. The latter is realised in our merger product as seen in Figure~\ref{figure:chemie}, hence, the strong He\,I line in SN\,2004dy might  favor the merger-CE origin of this SN.
On top of that, photospheric velocity in SN\,2004dy estimated via minimum of P\,Cyg of Fe\,II line are around 4,000~km\,s$^{\,-1}$ at day~30, which is close to the photospheric velocities of our merger at corresponding epoch (see Figure~\ref{figure:uph}). 

\subsection[Non-electromagnetic observables]{Non-electromagnetic observables}
\label{subsect:nonem}


In addition to the electromagnetic signal, multi-messenger observations could in principle lead to a clear distinction between a thermonuclear explosion and a core-collapse event. In particular the expected gravitational wave (GW) signal and neutrino radiation are expected to be fundamentally different.


It has recently been suggested that nearby CE events can produce a GW signal detectable by the upcoming space-based GW observatory LISA \footnote{\url{https://lisa.nasa.gov/}} and future planned deci-hertz observatories such as DECIGO \citep{2019IJMPD..2845001K} or BBO \citep{4453737} if they lead to a core merger \citep{Ginat2020,Moran-Fraile2023} or to a tight enough orbit \citep{renzo2021}.
The binary system studied in the present paper is similar to that considered in \citet{Moran-Fraile2023}, releasing GW radiation on a broad frequency range during the CE event ($f_\mathrm{GW}\sim10^{-5}~\mathrm{Hz}$) until the core merger ($f_\mathrm{GW}\sim10^{-2}~\mathrm{Hz}$), and should be detectable under the same conditions.
This signal will only diverge from the one presented in \citet{Moran-Fraile2023} in the post-merger stage, as the thermonuclear explosion taking place shortly after the disruption of the core will avoid the emission of most of the high-frequency components.



The GW signal of our merger will be fundamentally different to GW radiation produced by CCSNe, as the GW signal from CCSNe spreads in a very different frequency range between a few hundreds Hz and a few thousands Hz \citep[e.g., ][]{2017MNRAS.468.2032A,2023PhRvD.107j3015V}. Therefore, a multi-messenger detection would help identifying the explosion mechanism of such a transient. 

The event considered in our study would release neutrino radiation similar to other thermonuclear explosions \citep[e.g.,][]{2015PhRvD..92l4013S}.
In our simulations, the treatment of neutrinos is  very coarse and should be improved for a more detailed prediction of observables. It only takes into account thermal neutrino losses via a cooling term and it misses the neutrinos produced in weak reactions. At the point of carbon ignition, the merger releases $5\times10^{45}\,\mathrm{ergs}$ in neutrinos, which is much lower than typical total neutrino energy of $10^{\,53}$~ergs in neutrino-driven core-collapse explosions \citep{2010A&A...517A..80F,2016ApJ...818..124E,2021ApJ...909..169K}. Hence the lack of a detectable neutrino signal could be another diagnostic to distinguish between the thermonuclear and the core-collapse explosion scenarios, assuming that both explosions happen at close enough distances.



\section[Conclusions]{Summary and conclusions}
\label{sect:conclusions}

In the present study, we conducted radiative-transfer simulations for a merger of a white dwarf (WD) of 0.6~\Msun{} and a degenerate He-core of a red giant star \citep{Javier2023}. We show that the light curves of a pure merger are low-luminosity, as expected, because of low yield of radioactive nickel {}$^{56}$Ni (0.014~\Msun{}). 

We consider the system in which the merger occurs inside the atmosphere of a giant star, i.e. a binary undergoing the common-envelope (CE) episode. In this system, the CE is not fully ejected under certain conditions, but expands resulting in an extended envelope.  We consider two possibilities for CE extension, to the radius of 75~\Rsun{} and 330~\Rsun{}, without and with perturbations, correspondingly. These two possibilities can be interpreted as follows. The unperturbed envelope is a default atmosphere of a red giant star with an initial mass of 2~\Msun{} at the end of the hydrogen core burning, while the perturbed envelope is a result of an evolution with the additional injected energy which mimics the heat from the orbital energy coming from the inspiraling a WD and a He-core. 

A WD-He-core merger happening inside the hydrogen-rich atmosphere of a giant star will have observational properties different to a SN~Ia transient. Instead, the predicted transient LC resembles a short 30\,--\,45~day plateau, Type\,II SNe. Spectroscopically the transient is supposed to resemble a hydrogen-rich SN during the plateau phase, as the atmosphere of a giant star contains 1~\Msun{} of hydrogen. Depending on the viewing angle, the plateau lasts 30 to 45~days. Luminosity on the plateau is intermediate to low, $\log L_\mathrm{bol} \sim 41.7$ to 42~erg\,s$^{\,-1}$, i.e. $-15.4$~mags to $-16.4$~mags in $V$-band.

We found five SNe with the short 30--40~day plateau among the observed SNe~IIP available in the literature, which can be candidates for our models. These are: SN~2004dy, SN~2005af, SN~2005hd, SN~2007aa, and SN~2008bu. However, our analysis was limited because of a lack of a full observational set for these events. 

Our models at later phases will have spectra which are supposed to be a mixture of spectra both Type~Ia and Type~II SNe, and low photospheric velocities about 2,500~km\,s$^{\,-1}${}. Hence, late time, i.e. later than 40~days after the first detection, observations will help to distinguish events like models in our study.
The merger inside the CE will also have a certain GW signal, which is different to WD-WD background noise and very different to GW signal from core-collapse SNe, and very low-luminosity neutrino radiation different to the neutrino radiation released by core-collapse explosions.

It is expected that in some binary stellar systems entering CE evolution, the envelope ejection is not successful and a CE-merger event of the two stellar cores inside the envelope ensues. We predict that the resulting transient will resemble SNe~II although the physical origin is a thermonuclear explosion of the merging cores.
An observational identification of the transients resulting from such a scenario based on the predicted synthetic observables will shed light on the CE physics. 

\section*{Acknowledgments}

We thank Petr Baklanov, St{\'e}phane Blondin, Daniel Kresse, and Patrick Neunteufel for helpful discussions.

This work has been supported by the Klaus Tschira Foundation. 
J.M-F., A.H.\ and V.A.B.\ are fellows of the International Max Planck Research School for Astronomy and Cosmic Physics at Heidelberg (IMPRS-HD) and acknowledge financial support from IMPRS-HD.

This work was supported by the High Performance and Cloud Computing Group at the Zentrum f{\"u}r Datenverarbeitung of the University of T{\"u}bingen, the state of Baden-W{\"u}rttemberg through bwHPC and the German Research Foundation (DFG) through grant no INST 37/935-1 FUGG.

%

\bibliography{references}{}
\bibliographystyle{aa}


\appendix

\section[Broad-band LCs of the selected models: AA and X-ray cases]{Broad-band LCs of the selected models: AA and X-ray cases}
\label{appendix:append1}

\begin{figure*}
\centering
\includegraphics[width=0.5\textwidth]{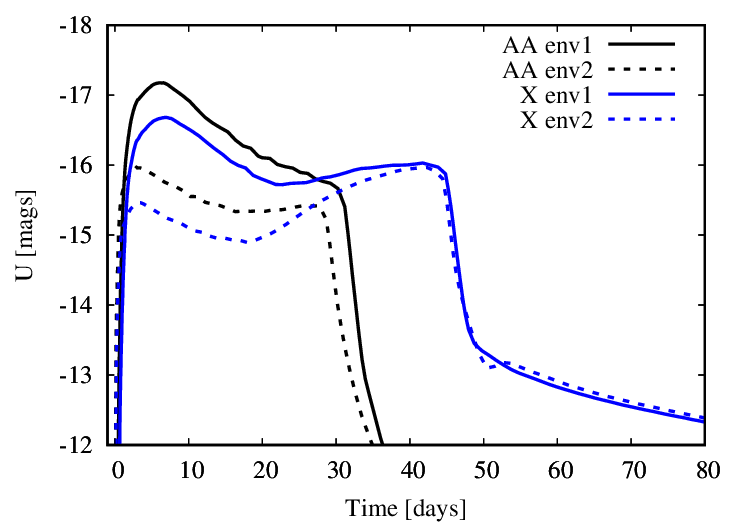}~
\includegraphics[width=0.5\textwidth]{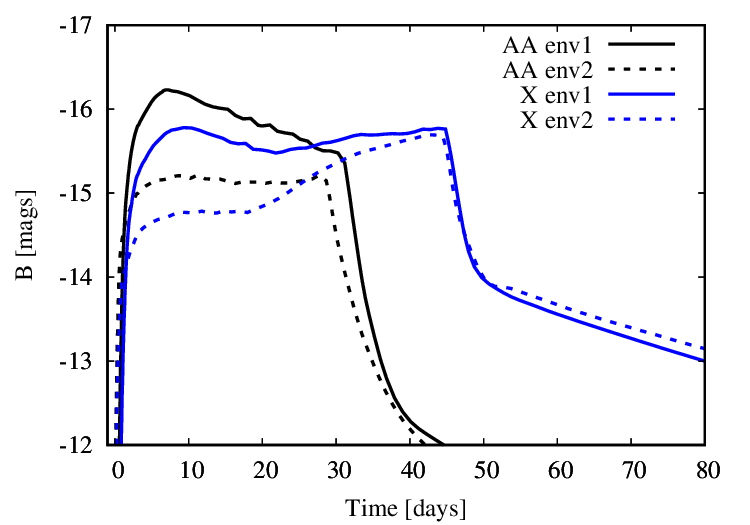}\\
\includegraphics[width=0.5\textwidth]{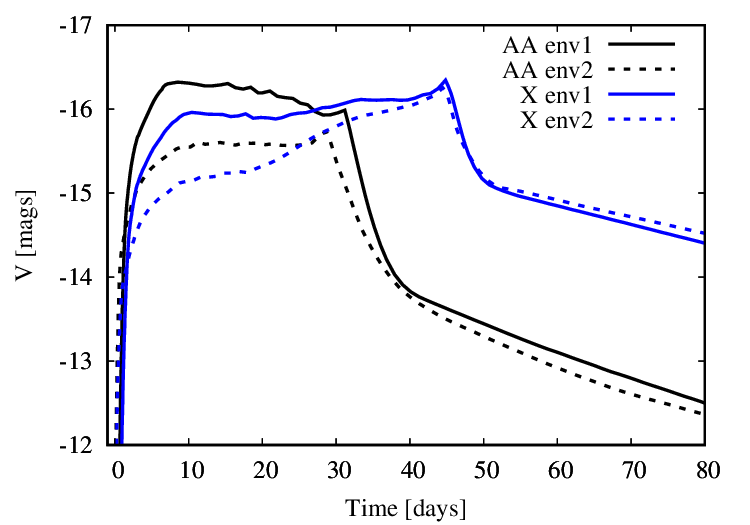}~
\includegraphics[width=0.5\textwidth]{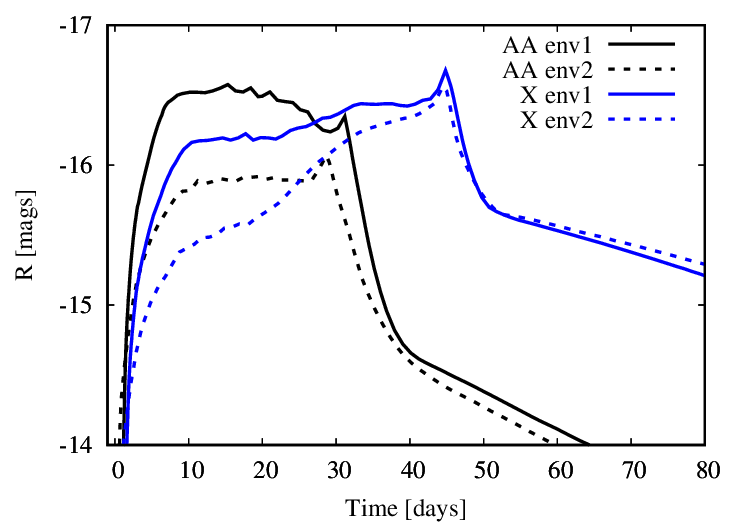}
\caption{$U$, $B$, $V$, and $R$ broad band LCs for the AA case (black) and the
$x$-ray (blue) in the extended ``env1'' (solid) and compact ``env2'' (dashed)
envelopes. The label ``X'' stands for the case $x$ in our set of selected directions.}
\label{figure:UBVRadd}
\end{figure*}

In Figure~\ref{figure:UBVRadd}, we present $U$, $B$, $V$, and $R$ broad band LCs
for the AA case (black) and the
$x$-ray (blue) in the extended ``env1'' (solid) and compact ``env2'' (dashed)
envelopes. These plots show that the models within the compact envelope have
intermediate to low luminosity during the plateau phase, $-15$~mags to $-16$~mags. This
low magnitude in combination with the very short plateau of about 30 days
(AA; black) and 45~days ($x$-ray; blue) makes it difficult to find a suitable candidates among observed SNe~IIP
\cite{2014ApJ...786...67A,2014MNRAS.442..844F,2022A&A...660A..40M}. However, we succeeded with this task and find five possible candidates as discussed in Section~\ref{sect:obs}. We conclude that the case of the merging WD-He-core system
within a compact envelopes is very rare in Nature. In our search, one SN out of five candidates is found. The case of the same system being embedded into an extended
envelope is more probable, and based on the current observational record, we conclude that at least some SNe~IIP with normal
to low luminosity with extremely short plateau of 30--40~days can be a
result of evolution of a binary system, in which a WD engulfed into the
CE merges with the He-core of a red giant and detonates.

\section[Dependence on metallicity]{Dependence on metallicity}
\label{appendix:append2}

\begin{figure*}
\centering
\includegraphics[width=0.5\textwidth]{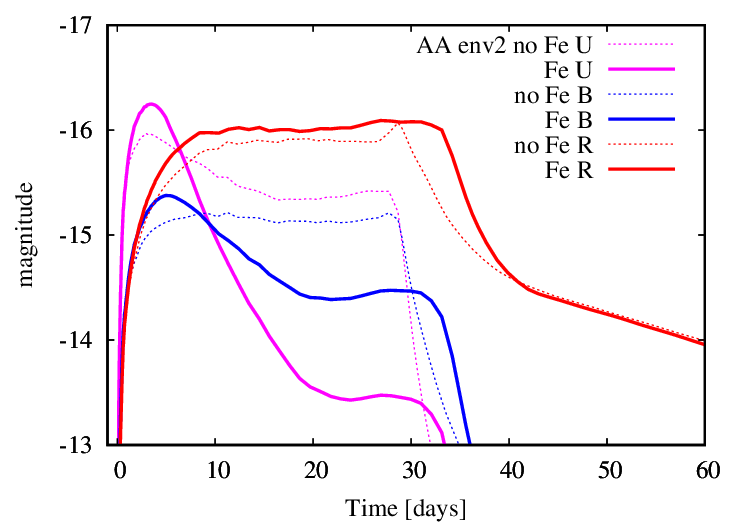}~
\includegraphics[width=0.5\textwidth]{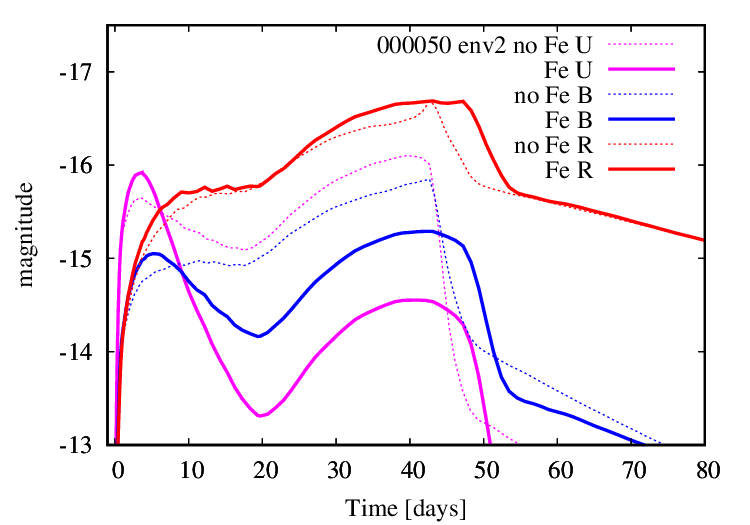}
\caption{Metallicity dependence for the compact envelope: 
$U$ (magenta), $B$ (blue), $R$ (red) LCs showing the difference
between the case of iron-free envelope (zero abundance of iron; dashed) and iron-polluted (iron abundance corresponding to solar metallicity; solid) envelope for the AA case (left) and the ray 000\,050 (right).}
\label{figure:ZdependAdd}
\end{figure*}

In Figure~\ref{figure:ZdependAdd}, we provide plots showing dependence on metallicity for the case
of the compact envelope of the giant star in addition to the problem
discussed in Section~\ref{subsect:Zdepend}.
The resulting LCs display similar behaviour as the case of the extended
envelope. Namely, the models with the iron polluted envelope with the iron fraction corresponding to the solar metallicity have $U$-band
LCs declining faster during the plateau phase, and end up with the
$U$-band magnitude 1~mags lower. The $R$-band LCs serve as a
representative of the bolometric LCs showing that the plateau lasts 5~days
longer than those LCs for the iron-free models. The latter is explained by the higher opacity in the hydrogen-rich
envelope, as the line opacity of the iron-group elements is the highest among all species.

\end{document}